# DHCAL with Minimal Absorber: Measurements with Positrons

## The CALICE Collaboration

**B. Freund[a], C. Neubüser[b], J. Repond[*], J. Schlereth, L. Xia**

*Argonne National Laboratory, 9700 S. Cass Avenue, Argonne, IL 60439, U.S.A.*

**A. Dotti[c], C. Grefe[d], V. Ivantchenko**

*CERN, 1211 Genève, 23, Switzerland*

**J. Berenguer Antequera, E. Calvo Alamillo, M.-C. Fouz, J. Marin, J. Puerta-Pelayo, A. Verdugo**

*CIEMAT, Centro de Investigaciones Energeticas, Medioambientales y Tecnologicas, Madrid, Spain*

**E. Brianne, A. Ebrahimi, K. Gadow, P. Göttlicher, C. Günter, O. Hartbrich, B. Hermberg, A. Irles, F. Krivan, K. Krüger, J. Kvasnicka, S. Lu, B. Lutz, V. Morgunov[e], A. Provenza, M. Reinecke, F. Sefkow, S. Schuwalow, H.L. Tran**

*DESY, Notkestrasse 85, D-22603 Hamburg, Germany*

**E. Garutti, S. Laurien, M. Matysek, M. Ramilli, S. Schroeder**

*Univ. Hamburg, Physics Department, Institut für Experimentalphysik, Luruper Chaussee 149, 22761 Hamburg, Germany*

**B. Bilki, E. Norbeck[e], D. Northacker, Y. Onel**

*University of Iowa, Dept. of Physics and Astronomy, 203 Van Allen Hall, Iowa City, IA 52242-1479, USA*

**J. Cvach, P. Gallus, M. Havranek, M. Janata, M. Kovalcuk, J. Kvasnicka, D. Lednicky, M. Marcisovsky, I. Polak, J. Popule, L. Tomasek, M. Tomasek, P. Sicho, J. Smolik, V. Vrba, J. Zalesak**

*Institute of Physics, Academy of Sciences of the Czech Republic, Na Slovance 2, CZ-18221 Prague 8, Czech Republic*

**B. van Doren, G. W. Wilson**

*University of Kansas, Department of Physics and Astronomy, Malott Hall, 1251 Wescoe Hall Drive, Lawrence, KS 66045-7582, USA*

**K. Kawagoe, Y. Miyazaki, Y. Sudo, T. Suehara, T. Tomita, H. Ueno, T. Yoshioka**

*Department of Physics, Kyushu University, Fukuoka 812-8581, Japan*

**S. Bilokin, J. Bonis, P. Cornebise, R. Pöschl, F. Richard, A. Thiebault, D. Zerwas**

*Laboratoire de L'accélerateur Linéaire, Centre d'Orsay, Université de Paris-Sud XI, BP 34, Bâtiment 200, F-91898 Orsay CEDEX, France*

**J. -Y. Hostachy, L. Morin**

*Laboratoire de Physique Subatomique et de Cosmologie - Université Grenoble-Alpes, CNRS/IN2P3, Grenoble, France*

**D. Besson, M. Chadeeva**[f] **, M. Danilov**[f,g] **, O. Markin**[h] **, E. Popova**

*National Research Nuclear University MEPhI (Moscow Engineering Physics Institute) 31, Kashirskoye shosse, 115409 Moscow, Russia*

**M. Gabriel, P. Goecke, C. Kiesling, N. van der Kolk, F. Simon, M. Szalay**

*Max Planck Inst. für Physik, Föhringer Ring 6, D-80805 Munich, Germany*

**F. Corriveau**

*McGill University, 3600 University Street, Montreal, QC H3A2T8, Canada*

**G. C. Blazey, A. Dyshkant, K. Francis, V. Zutshi**

*NICADD, Northern Illinois University, Department of Physics, DeKalb, IL 60115, USA*

**K. Kotera, H. Ono**[i] **, T. Takeshita**

*Shinshu Univ., Dept. of Physics, 3-1-1 Asaki, Matsumoto-shi, Nagano 390-861, Japan*

**S. Ieki, Y. Kamiya, W. Ootani, N. Shibata**

*ICEPP, The University of Tokyo, 7-3-1 Hongo, Bunkyo-ku, Tokyo 113-0033, Japan*

**D. Jeans, S. Komamiya, H. Nakanishi**

*Department of Physics, Graduate School of Science, The University of Tokyo, 7-3-1 Hongo, Bunkyo-ku, Tokyo 113-0033, Japan*

[a] *Also at McGill University, Montreal, Canada*
[b] *Also at DESY, Hamburg, Germany*
[c] *Now at SLAC, Stanford, CA, USA*
[d] *Now at Bonn University, Bonn, Germany*
[e] *Deceased*
[f] *Also at P. N. Lebedev Physical Institute, Russian Academy of Sciences*
[g] *Also at Moscow Institute of Physics and Technology (MIPT), Moscow, Russia*
[h] *Also at Institute of Theoretical and Experimental Physics, Moscow, Russia*
[i] *Now at Nippon Dental University, Niigata, Japan*

*Corresponding author, E-mail*: `repond@anl.gov`

ABSTRACT: In special tests, the active layers of the CALICE Digital Hadron Calorimeter prototype, the DHCAL, were exposed to low energy particle beams, without being interleaved by absorber plates. The thickness of each layer corresponded approximately to 0.29 radiation lengths or 0.034 nuclear interaction lengths, defined mostly by the copper and steel skins of the detector cassettes. This paper reports on measurements performed with this device in the Fermilab test beam with positrons in the energy range of 1 to 10 GeV. The measurements are compared to simulations based on GEANT4 and a standalone program to emulate the detailed response of the active elements.

# Contents

## 1. Introduction

Imaging calorimeters, i.e. calorimeters with very fine lateral and longitudinal segmentation of the readout, offer distinct advantages over the traditional equivalent with a tower structure. Among these are their enhanced capability for the application of Particle Flow Algorithms to the measurement of hadronic jets [1], software compensation techniques to improve the energy resolution of single particles [2], longitudinal leakage corrections for hadronic showers, and particle (electron, muon, hadron) identification algorithms.

The CALICE collaboration [3] is developing several different sensors as candidates for the active media of imaging calorimeters: Silicon pads [4], Scintillator pads [5] and strips [6], Resistive Plate Chambers (RPCs) [7,8], Micromegas [9], and Gas Electron Multipliers (GEMs) [10], where the gaseous devices are read out with small, square pads. This paper reports on special tests performed in the Fermilab test beam using the detector cassettes of the Digital Hadron Calorimeter prototype, the DHCAL [7], without absorber material interleaved between the active layers. The active layers of the DHCAL contained thin RPCs with a readout featuring $1 \times 1$ cm$^2$ pads.

In its configuration without absorber plates, the so-called Min-DHCAL provided the opportunity to study electromagnetic and hadronic showers with extremely fine segmentation, especially longitudinally, spreading the showers over the entire depth of the stack. This paper presents measurements with positrons in the energy range of 1 – 10 GeV. The experimental results are compared to detailed Monte Carlo simulations based on GEANT4. These comparisons provide an ideal tool to gain deeper insights into the physics of an RPC–based calorimeter, in particular at the low energies of the present measurements. Since the uncertainties in the simulation of electromagnetic showers are relatively small, at the percent level, the comparison can be used to tune the modeling of the DHCAL in the simulation. Once tuned, the model can be used to provide absolute predictions for the response to hadrons, which in turn can be compared to experimental measurements. The comparisons of the hadronic response will be the subject of a forthcoming paper.

## 2. Description of the DHCAL with Minimal Absorber

The DHCAL used Resistive Plate Chambers [11] as active elements. The area of each RPC measured $32 \times 96$ cm$^2$. The chambers utilized the traditional two resistive-plate design with soda-lime glass as resistive plates [12]. The cathode and anode plates were 1.15 and 0.85 mm thick, respectively, and enclosed a single 1.15 mm thick gas gap. The chambers were flushed with a non-flammable mixture of three gases: tetrafluoroethane (94.5%), isobutane (5.0%) and sulfur hexafluoride (0.5%) and were operated in avalanche mode with a default high voltage of 6.3 kV.

The readout boards each measured $32 \times 48$ cm$^2$, contained 1,536 $1 \times 1$ cm$^2$ pads, and were placed on the anode side of the chambers. Two boards covered the area of one chamber. The readout boards contained two separate boards, a board featuring the pads and a board housing the front-end electronics, interconnected by dots of conductive glue. The electronic readout system was based on the DCAL III chip [13], which applied a single threshold to the signals from an array of $8 \times 8$ readout pads to define hits [14]. This type of electronic readout with single-bit resolution per channel is commonly referred to as 'digital'. The threshold discriminating the signals could be set individually for each chip, but was common to all 64 channels connected to a chip. The readout was pulsed with a 10 MHz clock. After receipt of an external trigger, hit patterns of the 64 channels connected to a chip and in seven consecutive 100 ns time bins were read out together with their corresponding time stamps. Of the seven time bins, one bin corresponded to times before the arrival of the particle in the stack.

Three RPCs and their six corresponding readout boards were assembled into a cassette to serve as an active layer of the calorimeter. Thus the active area of a cassette was approximately $96 \times 96$ cm$^2$. The cassettes featured a front-plate (2 mm copper) and a rear plate (2 mm steel) held together by rectangular bars located on the top and bottom of the structure. The thickness of each cassette was about 12.5 mm and corresponded to 0.29 radiation lengths ($X_0$) or 0.034 nuclear interaction lengths ($\lambda_I$). Slight variations of the effective thickness across the surface of the cassettes, due to the packaged DCAL III chips, have been ignored in the simulation of the set-up.

The Min-DHCAL consisted of 50 cassettes, spaced every 2.54 cm. The thickness of the entire stack corresponded to approximately 15 $X_0$ or 1.7 $\lambda_I$. The total number of readout channels was 460,800, which at the time constituted a world record for calorimetry in High Energy Physics. Figure 1 shows a photograph of the Min-DHCAL.

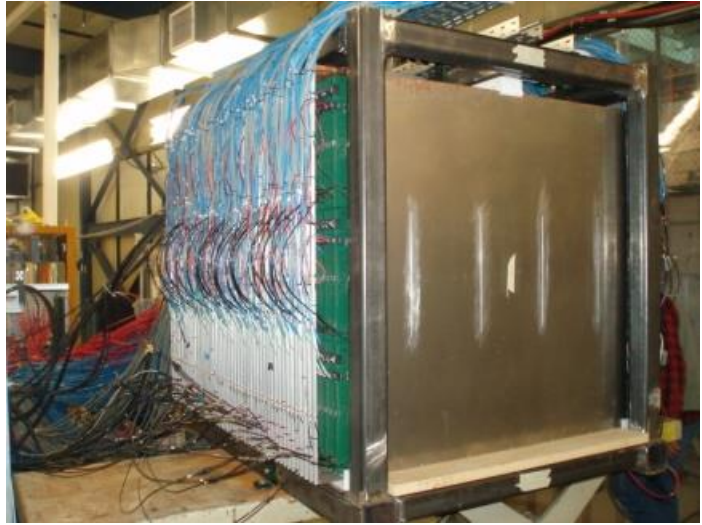

**Figure 1.** Photograph of the DHCAL with minimal absorber (the Min-DHCAL) in the Fermilab test beam.

With the single-bit readout, the energy of an incident particle is reconstructed in the DHCAL to first order as the number of pads hit, i.e. pads with a signal above threshold. This method works, because the accidental noise rate in the stack was extremely small. The average number of accidental noise hits was measured to be ~0.2 in a reconstructed event (after applying all the cleaning cuts of the standard analysis, see below). Assuming a calibration of 90 hits per GeV, as was observed for 6 GeV positrons (see below), the average noise rate per event corresponds to 2.3 MeV and is therefore negligible.

## 3. Data Collected at the Fermilab Test Beam Facility

The Min-DHCAL was exposed to the test beam at the Fermilab Test Beam Facility, FTBF [15]. The facility provides a primary 120 GeV proton beam and momentum selected secondary beams in the range of 1 – 66 GeV/c (energy and momentum units are used interchangeably throughout the text). The latter are a mixture of electrons, muons and pions, where the fraction of electrons is dominant at momenta below 5 GeV/c and tapers off for momenta above 32 GeV/c. The particles arrive every 60 seconds in spills of 4.0 seconds duration. The beamline included two Čerenkov counters for particle identification and two scintillator paddles (19 × 19 $cm^2$), located approximately two meters upstream of the Min-DHCAL. The data acquisition was triggered by the coincidence of these two paddles. The readout system time stamped the trigger signal and recorded this information into the event header. Of the two available Čerenkov counters, only one was used in these tests. Its discriminated signal was read into the data stream and used offline to separate positrons from muons and pions.

The data on which this paper is based were collected in November 2011. Runs were taken with a selected momentum in the range of 1 - 10 GeV/c, as seen in Table I.

Table I. Summary of the data taken in the Fermilab test beam with the Min-DHCAL.

| Momentum [GeV/c] | Number of events |
|---|---|
| 1 | 107k |
| 2 | 117k |
| 3 | 62k |
| 4 | 84k |
| 6 | 109k |
| 8 | 109k |
| 10 | 226k |
| **Total** | **814k** |

### 4. Simulation of the Test Beam Set-up

The simulation of the test beam set-up is based on the GEANT4 program [16] version 10.02, with default parameters. The simulated set-up includes the active elements with their cassette covers, resistive plates, gas gap, and electronic readout boards, in addition to the material in the beam line upstream of the stack, corresponding to 0.2 $X_0$. The generated momenta of the particles were smeared by 2.5%, reflecting the momentum spread of the FTBF test beam [15].
For each momentum setting, the beam profile, as measured by the Min-DHCAL, was emulated by randomly smearing the positrons' impact points on the calorimeter and their angles of incidence.

The mean ionization energy and the so-called Fano factor for the default RPC gas mixture were calculated with HEED [17] and amount to 33.25 eV and 19, respectively. These values are provided externally to GEANT4, as the mixture has not yet been defined within the GEANT4 package. Any energy deposition generated by the simulation in the gas gap of the RPCs is used as a seed for creating an avalanche. The procedure ignores the actual amount of energy deposited by a particle in the gas gap, as it is only weakly correlated to the signal charge.

The response of the RPCs is simulated using a standalone program, called RPC_sim [18]. RPC_sim generates a charge according to measurements of the avalanche charge distribution [12], spreads the charge onto the anode plane and defines hits in pads by applying a threshold on the charge. The generated charge is spread over the anode plane assuming a drop-off with distance from the center of the avalanche described by the sum of two one-dimensional Gaussians. The charges induced by different avalanches on a given pad are summed up. A threshold is applied on the summed-up charge to define a hit. The simulation of the RPC response is governed by six parameters, which once determined are fixed for all types of particles and all energies. Of these, five are tuned by comparing the distribution of the number of hits per layer in both measured and simulated muon track events. These parameters include the widths of the Gaussians, their relative weight, the charge threshold to define hits and a parameter to fine-adjust the generated charge distribution. The latter takes into account differences in both chamber construction and operation compared to the analog measurements of [12]. The remaining parameter, $d_{cut}$, a distance cut in the $x/y$ plane[1] introduced to suppress close-by avalanches, is tuned using showers with more than a single avalanche per plane. Here the 3 and 10 GeV positron data are used to determine its value. The $d_{cut}$ parameter is tuned by comparing the measured and simulated number of hits (both the peak position and the width of the distribution), the longitudinal shower shape, and the distribution of the density of hits (see section 9). The optimized parameter minimizes the sum of the $\chi^2$ of the differences between the measured and simulated distributions.

Initially, the GEANT4 program utilized the FTFP_BERT physics list. However, this led to a unsatisfactory description of, among other measurements, the energy resolution, suggesting the generation of too few initial ionizations in the gas gap of the RPCs. A migration to the 'Option 3' or '_EMY' based electromagnetic physics list, which is particularly appropriate for low energies, resulted in a significant improvement of the description of the experimental data. In the following, unless noted otherwise, the results of the simulation are based on FTFP_BERT_EMY physics list.

## 5. Hit and Event Selection

Between data taking runs and with the beam off, data were collected in trigger-less mode to monitor the status of the chambers and the data acquisition system. These so-called 'Noise runs' typically lasted 60 seconds and provided an important and quick cross-check of the status of the Min-DHCAL. These runs revealed that for a large number of chambers the pads close to the ground lead on the resistive layer tend to fire randomly at a relatively elevated rate. In order to reduce the contamination from accidental hits due to this local high rate, any hits in an area of 2 $\times$ 5 cm$^2$ around the ground lead were ignored. The same cut was applied to the simulated data, resulting in a negligible loss of hits.

For a selected data run collected with a beam setting corresponding to 10 GeV/c, Fig. 2 shows the distribution of the time differences between time-stamps of the trigger and of the corresponding hits in an event. Most hits occur with a difference of 19 and 20 time bins, where each bin corresponds to 100 ns. In the following, only events with most of their hits in either of these two time bins were analyzed, thus removing around 0.2 % of the events. A manual scan of the events discarded showed that this cut effectively reduced the number of events with multiple particles, often spread over different time-bins. In addition to the above requirements, any hit in an analyzed event with a time stamp difference not equal to 19 and 20 was ignored. This cut typically removed (1 – 2) % of the hits and helped reduce the contribution from accidental noise hits.

---

[1] The coordinate system is defined such that $x$ is horizontal and perpendicular to the beam, $y$ is vertical, and $z$ is in the beam direction. The axes correspond to a right-handed coordinate system.

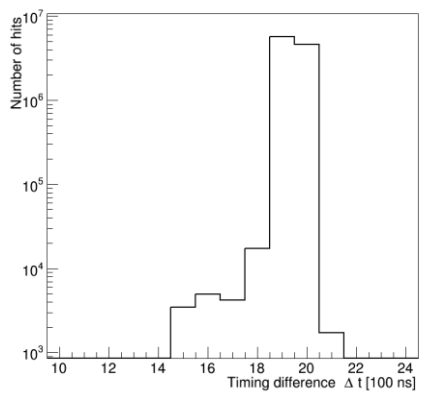

**Figure 2.** Distribution of the time difference between time-stamps of the trigger and the corresponding hits for events in a selected 10 GeV/c run. Each bin corresponds to 100 ns and earlier (later) hits lead to larger (smaller) time differences.

A small fraction of the DCAL chips (<1%) were dead and therefore provided no hits. The data were not corrected for this inefficiency, but rather hits in the areas read out by the dead chips were also discarded in the simulation.

Finally, a small fraction of hits in the same event showed the same geometrical address, but occurred with different time-stamps. These duplicate hits were only counted once.

The hits in each layer of the DHCAL were clustered using a simple nearest-neighbor clustering algorithm. Two hits were assigned to the same cluster if they shared a common side. Events containing multiple particles and showers which initiated upstream of the DHCAL were suppressed by requiring a single cluster of hits with a maximum of four hits in the first layer of the stack. This event quality selection eliminated between (10 − 20) % of the data, consistent with the fraction of events with multiple particles in the beam and/or early showers [15]. Of the simulated positron events, <3% are eliminated by this cut. Events acquired with spurious triggers (e.g. cosmic rays crossing the trigger counters) were eliminated with a requirement of hits in at least six different layers of the stack.

After the above selection, the data contain a mixture of positrons, muons, and pions. By requiring a signal in the upstream Čerenkov counter, the remaining fraction of muon and pion induced events was effectively reduced to zero, due to its negligible rate of accidental hits. The Čerenkov requirement was not applied to the simulated data. Table II summarizes the number of events passing each event selection requirement for both data and MC generated events as function of momentum. As can be seen from the entries for the simulated positron events, only a small fraction of at most 3.2 % of the events are discarded by the selection criteria.

## 6. Equalization of the RPC Responses

Through-going muon tracks are used to measure and equalize the response of the 150 different RPCs in the stack. The equalization procedure is applied for each data taking run individually to account for variations in the operating conditions (temperature and ambient air pressure). Muon tracks are selected by requiring at least one hit in both the first three and the last three layers of the stack. To reduce the number of interacting muons no two consecutive layers are allowed to feature four or more hits. In addition, none of the layers may contain more than one cluster of hits. The slope of the track is limited to at most 0.5 pads per layer, both horizontally and vertically. Typically, 5 – 10% of the events in a given run survive these selection cuts and are thus identified as through-going muons.

**Table II.** Cumulative percentage of events surviving the various event selection criteria. The Čerenkov requirement effectively eliminated contamination from muons/pions in the data, but was not applied to the simulated positron events.

| Data | Momentum [GeV/c] | 1 | 2 | 3 | 4 | 6 | 8 | 10 |
|---|---|---|---|---|---|---|---|---|
| | Timing cuts | 99.9 | 99.8 | 99.9 | 99.8 | 99.95 | 99.95 | 99.96 |
| | Requirements on first layer | 88.5 | 87.0 | 80.3 | 80.3 | 88.1 | 86.6 | 88.2 |
| | At least 6 active layers | 88.1 | 86.4 | 80.0 | 79.8 | 88.0 | 86.5 | 88.1 |
| | Čerenkov signal | 60.3 | 31.7 | 40.0 | 30.7 | 53.9 | 41.7 | 33.0 |
| Simulation | Timing cuts | 100.0 | 100.0 | 100.0 | 100.0 | 100.0 | 100.0 | 100.0 |
| | Requirements on first layer | 98.3 | 97.9 | 97.9 | 97.6 | 97.2 | 97.1 | 96.8 |
| | At least 6 active layers | 98.3 | 97.9 | 97.9 | 97.6 | 97.2 | 97.1 | 96.8 |

The efficiency $\varepsilon$ for detecting a minimum ionizing particle is calculated as the ratio of tracks producing at least one hit to the total number of tracks crossing a given chamber. The average pad multiplicity $\mu$ is determined as the average number of hits for tracks which generate at least one hit in that chamber. The calibration factor $c_i$ for chamber $i$ is then calculated as the product of the efficiency and pad multiplicity averaged over the entire stack and run period $\varepsilon_0\mu_0$, divided by the same product $\varepsilon_i\mu_i$, as determined for chamber $i$ in a given run:

$$c_i = \frac{\varepsilon_0 \mu_0}{\varepsilon_i \mu_i} \qquad (1)$$

The average efficiency $\varepsilon_0$ (pad multiplicity $\mu_0$) of the Min-DHCAL was determined to be 91.7% (1.573).

The corrected number of hits, $N_i'$, is obtained by multiplying the measured number of hits, $N_i$, in chamber $i$ by its calibration factor

$$N_i' = c_i N_i \qquad (2)$$

As an illustration of the effect of the equalization procedure, Fig. 3(left) shows the peak position of the number of hits in the calorimeter for through-going muons, both before and after equalization as a function of run number. Through the equalization procedure the scattering of the points is shown to decrease significantly. When fit to a constant, the equalization procedure reduces the reduced $\chi^2$ of the fit from 24.2 to 0.8. However, the current method is seen to not be perfect, leaving some remaining differences between runs. This is also evident from Fig. 3(right), which shows the peak positions of the hit distributions for 10 GeV positrons versus run number. Various, more sophisticated, equalization procedures are currently under study and will be presented in a forthcoming paper. In simulated events, the dependence of the response as function of muon momentum is seen to be negligible in the energy range of the present study.

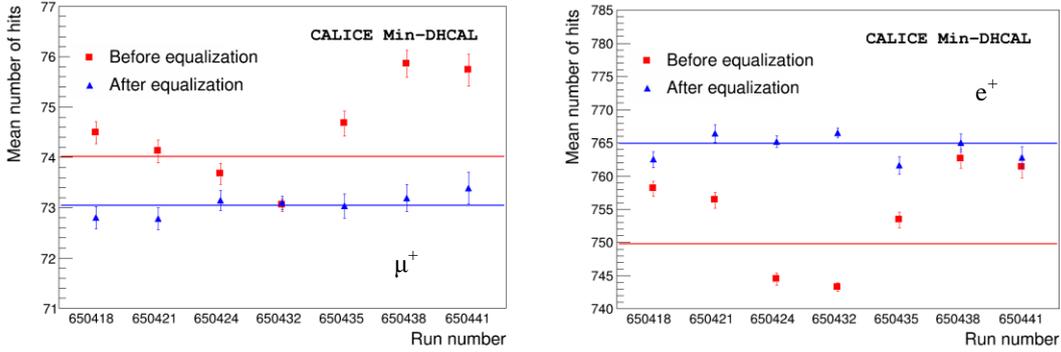

**Figure 3.** Peak position of the number of hits for through-going muon tracks (left) and positrons (right) collected at 10 GeV/c as function of run number, both before and after equalization of the response of the individual RPCs.

## 7. Systematic Errors

This section discusses the systematic errors associated with the experimental measurements as well as with the Monte Carlo simulation of the set-up. The following experimental systematic uncertainties were considered:

- To account for residual non-uniformity in the calibrated response of the RPCs in the stack, a systematic error of 50 % of the average difference between the results obtained before and after equalization has been assigned to each measurement. Defined this way, this systematic error amounts typically between 1 – 4 % of a given measured quantity. For most measurements this error is applied in both directions (positive and negative). However, for the resolution the error is only applied in the negative direction, since calibration errors can only degrade the resolution, but not improve it.

- For high incident particle rates, RPCs can demonstrate a loss of efficiency [19]. When exposed to the FTBF test beam, the chambers are able to recover their full efficiency in the ~56 second time gap between spills. However, during the spill the efficiency has been measured to first drop exponentially and then to remain constant at a lower value [19]. In order to estimate this effect on the present measurements, for every event its time with respect to the beginning of a spill was determined. An additional cut accepting only events in the first 0.5 second of a spill typically resulted in differences of 1 – 2 % in the number of hits compared to the default selection which accepts all events in a spill. Since this effect is small, no correction was applied, but rather a systematic error corresponding to this difference was assigned to all measurements. Since a loss of efficiency can only result in a decrease of hits, the error is only applied in the positive direction.
- The responses to both muons and pions result in distributions of number of hits distinctly different from those obtained with positrons, with different widths and different mean values. However, no enhancements of events are observed at the positions where contamination from muons or pions might create an excess of events. The contamination from muons and pions in the positron sample was therefore estimated to be significantly less than 1 %.
- The contribution from accidental noise hits is estimated to be negligible, as in average the noise rate was estimated to correspond to 0.2 hits per event in the entire stack.

All systematic errors are assumed to be independent and are therefore added in quadrature. Since the dominant contributions are related to the equalization procedure, possible correlations between energy points are assumed to be negligible.

The differences observed between the FTFP_BERT and FTFP_BERT_EMY physics lists, see below, point to some uncertainty in the simulation of electromagnetic showers in GEANT4. Additionally, uncertainties in the emulation of the RPC response lead to further systematic errors in the simulated results. However, since the 3 and 10 GeV measurements were used to tune the distance cut parameter of the RPC_sim program, $d_{cut}$, the simulation lost most of its predictive power and an assignment of systematic errors to the simulation of positrons has therefore become problematic.

## 8. Response of the Min-DHCAL to Positrons

Figure 4 shows the distribution of the number of hits for all selected positron events for both data and simulation. The response curves have been normalized to unity for each momentum selection and are well described by fits of a Gaussian function in the range of ±2σ around the peak value (determined iteratively). At the level of $10^{-3}$ and below, the curves show tails, mostly towards lower number of hits. These tails are also present, but somewhat smaller in the simulation, and are an artefact of the digital readout of electromagnetic showers. The tails in the data might include contamination from particles other than positrons (<1%).

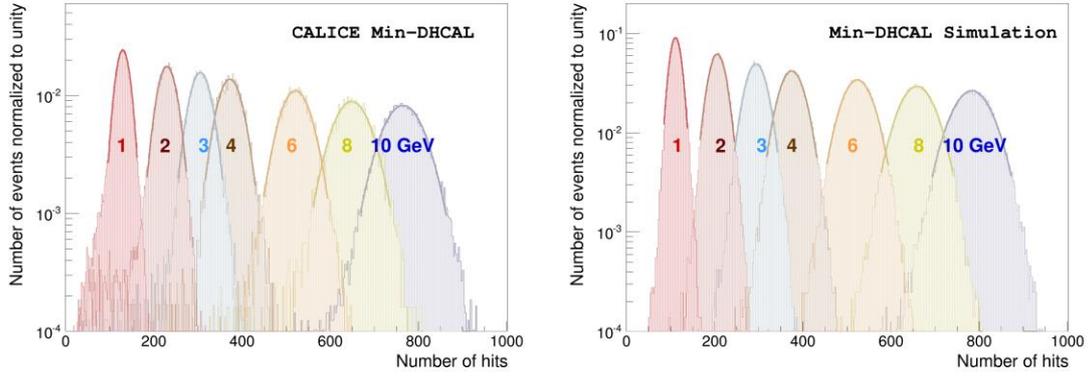

**Figure 4.** Distribution of the number of hits for all selected positron events for data (left) and simulation based on the FTFP_BERT_EMY physics list (right). The distribution is plotted separately for each beam momentum setting (1, 2, 3, 4, 6, 8, 10 GeV/c) and is normalized to unity. The distributions are fit with a Gaussian function in the range of ±2 standard deviations. The results of the fits are shown as solid lines.

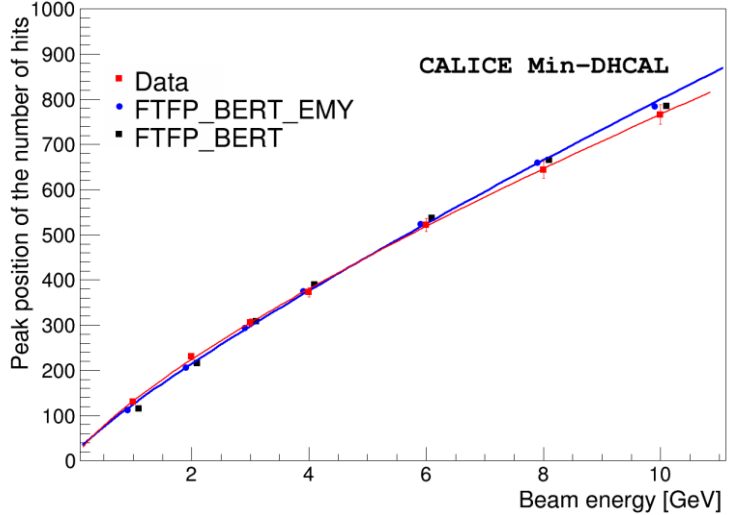

**Figure 5.** Peak position of the number of hits versus positron beam energy for both data (red) and simulation based on FTFP_BERT (black) and FTFP_BERT_EMY (blue). The experimental data points and the simulation based on FTFP_BERT_EMY have been fitted to a power law shown as solid lines. The error bars of the data include both the statistical and systematic (dominant) errors. The statistical error bars of the simulation are smaller than the marker size. The simulated points have been shifted in x for better visibility.

The mean values obtained from the Gaussian fits are shown as function of beam energy in Fig. 5. The statistical error is smaller than the size of the points. The systematic error is dominated by contributions from the calibration uncertainty. The data are compared to the results of the Monte Carlo simulation based on both the FTFP_BERT and the FTFP_BERT_EMY physics lists. Both are seen to be in good agreement with the data. The data/simulation are fit to a power law

$$N_{hit} = a_0 (E_{beam}/\text{GeV})^m \qquad (3)$$

where the exponent *m* is a measure of the non-linearity (saturation) of the response. A value of unity would indicate a perfectly linear response. A value of $m = 0.76 \pm 0.02$ $(0.836 \pm 0.001)$ is obtained for data (simulation based on FTFP_BERT_EMY), indicating a strong saturation of the response. The saturation is mostly due to the large pad size compared to the density of particles in the core of electromagnetic showers. The observed difference between data and simulation is due to a trend of the simulation to feature less hits at low energy and more hits at higher energy compared to the measurements. The values for the scale parameter $a_0$ for data and simulation (131.8 ± 2.9) and (115.8 ± 0.1), respectively. The simulation based on the FTFP_BERT physics list also produces similar results as the ones based on FTFP_BERT_EMY, as indicated by the black squares in Fig. 5.

The inverse of the power law is utilized to reconstruct the energy of the positrons. Figure 6 shows the reconstructed energy spectra for both data and simulation, again normalized to unity for each beam setting. The distributions were fitted to a Gaussian function in the range of ±2 standard deviations (determined iteratively). Figure 7 shows the resulting widths as function of beam energy for both data and simulation (based on both physics lists). The measured widths are approximately 15% better than the corresponding resolutions obtained by the simulation based on the FTFP_BERT physics list, indicating a possible deficit in the number of ionizations in the gas gap. On the other hand, the simulation based on the FTFP_BERT_EMY physics list reproduces the measurements quite well, but are in average about 6% better than the data. The energy resolution versus beam energy was fitted to the standard parametrization with a constant and a stochastic term

$$\frac{\sigma_E}{E} = c \oplus \frac{\alpha}{\sqrt{E/\text{GeV}}} \qquad (4)$$

In the following, only the simulated results based on the FTFP_BERT_EMY physics list will be shown.

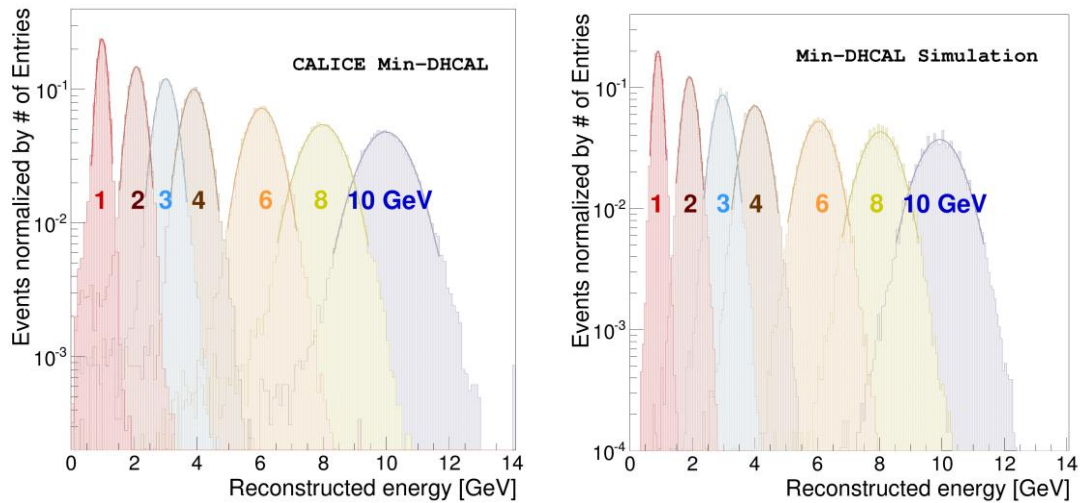

**Figure 6.** Reconstructed energy spectra for positrons: data (left) and simulation based on FTFP_BERT_EMY (right). The different beam momentum settings (1, 2, 3, 4, 6, 8, 10 GeV/c) are

– 11 –

indicated with different colors. The distribution at each beam setting was normalized to unity and was fit to a Gaussian function in the range of ±2 standard deviations.

Table III summarizes the results of the fits, showing a reasonable agreement between the stochastic terms of data and simulation. The comparatively large constant term in both data and simulation is most likely due to the above mentioned saturation effects, as the longitudinal leakage is relatively small (see below).

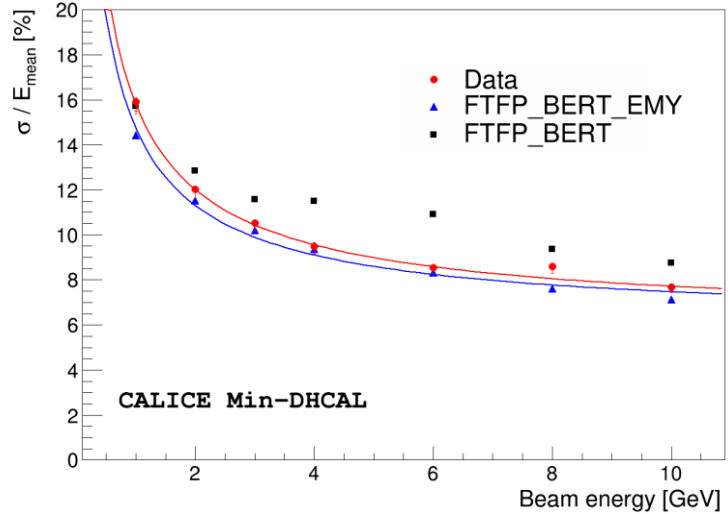

**Figure 7.** Energy resolution versus positron beam energy for data (red) and simulation based on FTFP_BERT (black) and FTFP_BERT_EMY (blue). The experimental points were corrected for the known momentum spread of the beam. The error bars of the data include both statistical and systematic uncertainties. The error bars of the simulation indicate the statistical uncertainty only. The curves are the results of fits to the quadratic sum of a constant and stochastic term, see Table III.

**Table III.** Fit parameters for the constant and stochastic terms of the energy resolution for positrons.

|  | $c$ [%] | $\alpha$ [%] |
|---|---|---|
| Data | 6.3±0.2 | 14.3±0.4 |
| Simulation (FTFP_BERT_EMY) | 6.2±0.1 | 13.4±0.2 |

## 9. Measurement of Shower Shapes

The imaging capabilities of the DHCAL provide an unprecedented tool for the detailed study of the shape of showers. The measurements in the present configuration with minimal absorber material spread electromagnetic showers over the entire depth of the Min-DHCAL stack, as seen in the event picture of Fig. 8.

As an example of the longitudinal shower shape, Fig. 9 shows the average shape measured with 6 GeV positrons. The measurement and the result of the simulation (histogram) show reasonable agreement. The shower maximum is observed around layer 20, where the data exhibit a slight deficiency compared to the simulation. This discrepancy does not seem to originate from the limited rate capability of the RPCs, as selecting events in the first half second of each spill result in the same depletion at shower maximum. The difference is most likely due to inaccuracies in the simulation of the RPC response and in particular of the spread of charges in the anode plane, but could also stem in part from the non-perfect equalization procedure of the RPC response.

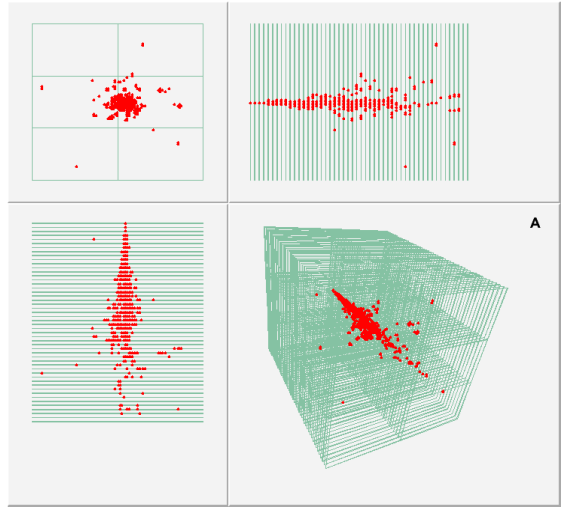

**Figure 8.** Display of an electromagnetic shower measured in the DHCAL and initiated by an 8 GeV positron. Starting from the upper left and going clockwise the views are: *x/y*, *x/z*, *x/y/z*, and *y/z*.

The longitudinal shapes are fit with a Gamma distribution [20], originally proposed for the description of the longitudinal energy deposition by electromagnetic showers, but here utilized to describe the number of hits as a function of layer number $z$

$$\frac{dN'}{dz} = N_0 b \frac{(bz)^{a-1} e^{-bz}}{\Gamma(a)} \qquad (5)$$

where $N_0$, $a$ and $b$ are free parameters. The result of the fit is shown as a solid line in Fig.9. In general, the fit describes the data well; however, the fit is not able to reproduce the shape around the shower maximum, where the fitted curve undershoots the measured points for both data and simulated events (not shown). Nevertheless, the location of the shower maximum can be obtained from the parameters $a$ and $b$ as

$$z_{max} = \frac{a-1}{b} \qquad (6)$$

Figure 10 shows the values of $z_{max}$ versus beam energy for both data and simulation. The experimental errors are dominated by systematic errors, but are smaller than the data points of the figure. The agreement between data and simulation is excellent.

The longitudinal dispersion is calculated for each event as

$$D_z = \sqrt{\frac{\sum z_i^2}{N} - \left(\frac{\sum z_i}{N}\right)^2} \qquad (7)$$

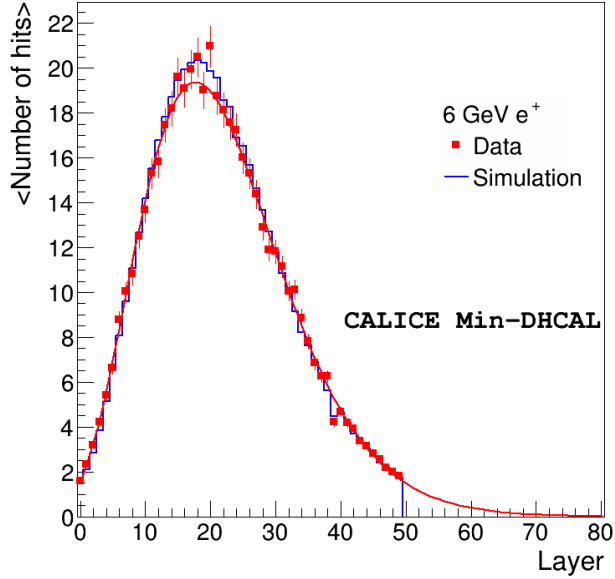

**Figure 9.** Measurement of the longitudinal shower shape for 6 GeV positrons, i.e. the calibrated average number of hits as function of layer number. The measurement (red points) is compared to the result of the simulation (blue histogram). The error bars of the data include both the statistical and systematic (dominant) uncertainties. The red line is a fit of the data with the Gamma distribution (see text).

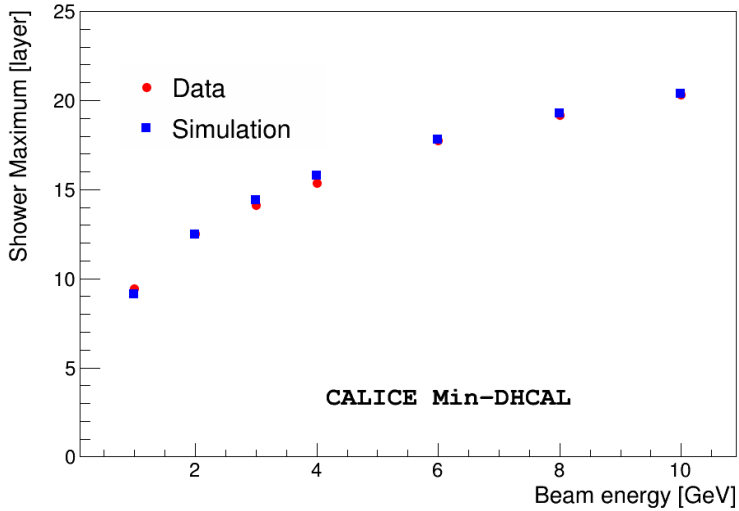

**Figure 10.** Average position of the shower maximum of positron induced events versus beam energy for both data (red) and simulation (blue). The error bars including statistical and systematic uncertainties for the data and statistical uncertainties only for the simulation are smaller than the marker size.

where the sum is over all hits and $N$ is the total number of hits in an event. The average longitudinal dispersion is plotted in Fig. 11 versus beam energy. The agreement between data and simulation is satisfactory with the data showing a slightly larger dispersion.

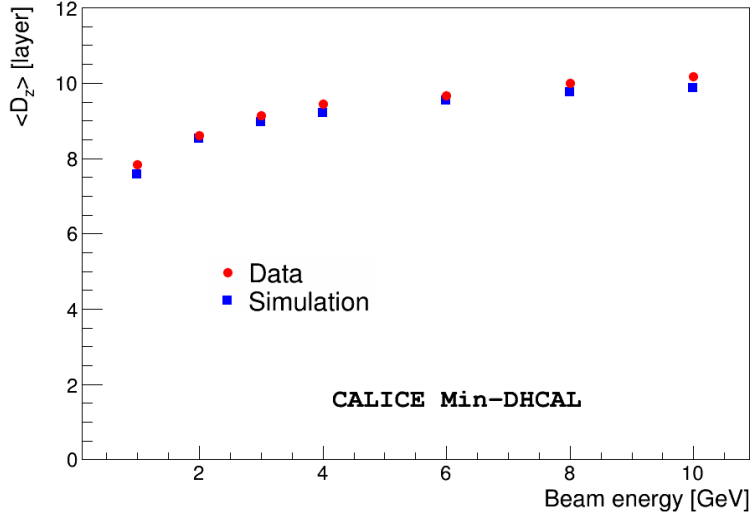

**Figure 11.** Dispersion of hits in longitudinal direction for positron induced events: data (red) and simulation (blue). The dispersion is averaged over all events at a given beam energy and is plotted as function of beam energy. The error bars including statistical and systematic uncertainties for the data and statistical uncertainties only for the simulation are smaller than the marker size.

In order to measure the transverse shower shape and to determine the direction of the incoming particle, for each event separate linear fits are performed to the hits in the $x/z$ and $y/z$ planes of the first five layers. In order to ensure a reliable reconstruction of the direction of the incoming particle, the following requirements are imposed: a) at least three layers among the first five feature hits, b) the position uncertainty for each hit in $x$ and $y$ is (arbitrarily) assumed to be $\pm 1$ cm. With this uncertainty, the $\chi^2$ of each fit be smaller than 1, and c) the reconstructed axis deviates by less than 0.1 radians from the average beam axis in both horizontal and vertical direction. As an example of the transverse shower shape, Fig. 12 shows the radial distance of each hit to the fitted straight line intersecting the corresponding detector plane, as measured for 6 GeV positrons. The accelerated decrease in entries above a radius of 50 cm is an artefact of the square shape of the detector planes with dimensions of $96 \times 96$ cm$^2$. Excellent agreement between data and simulation is observed over the entire range of radii apart from a small depletion at small radii in the data. Note that the number of hits varies over six orders of magnitude over the entire range in radii. Both the statistical and systematic uncertainties of the data are very small and mostly invisible in the plots.

Figure 13 shows the radial distance *R*, as measured using the method described above and averaged over all hits at a given beam energy. The average radial distance is particularly sensitive to the tails at high radii and is seen to be larger in data than simulation, notably at lower energies. The effect of additional noise in the data contributes insignificantly to this result, which was tested by adding noise hits to the simulated events. The noise hits were obtained from events collected with random triggers and no beam.

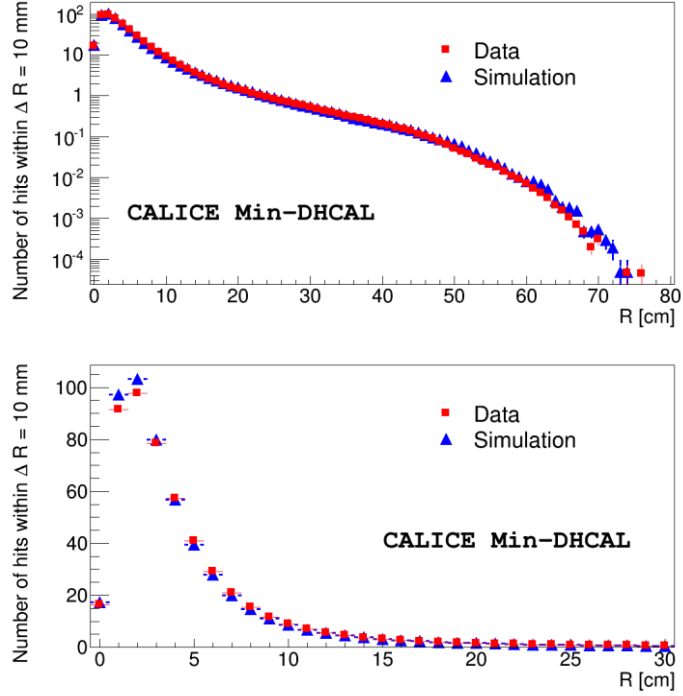

**Figure 12.** Distribution of the radial distance of hits from shower axis for 6 GeV positrons. The upper (lower) plot uses a logarithm (linear) y-scale. The areas of both plots are normalized to one event. The error bars include the statistical and systematic uncertainties for the data and statistical uncertainties only for the simulation.

The radial dispersion of hits in an event $D_r$ is calculated in a similar way to the longitudinal dispersion:

$$D_r = \sqrt{\frac{R_i^2}{N} - \left(\frac{\sum R_i}{N}\right)^2} \qquad (8)$$

Figure 14 shows the average radial dispersion as function of beam energy. Again, for lower energies the data show larger values compared to the simulation.

The density of hits is defined for each hit in an event as the number of hits in a volume of 3 × 3 × 3 pads surrounding the hit and can range from 0 to 26. Figure 15 shows the distribution of the density of hits for 6 GeV positrons for both data and simulation. The simulation shows a higher

– 16 –

probability for high hit densities than the data. In the simulation the hit density distribution was seen to depend on the value of the distance cut parameter, used to suppress close-by avalanches (see Section 4). The value providing the best agreement with the data was chosen as default.

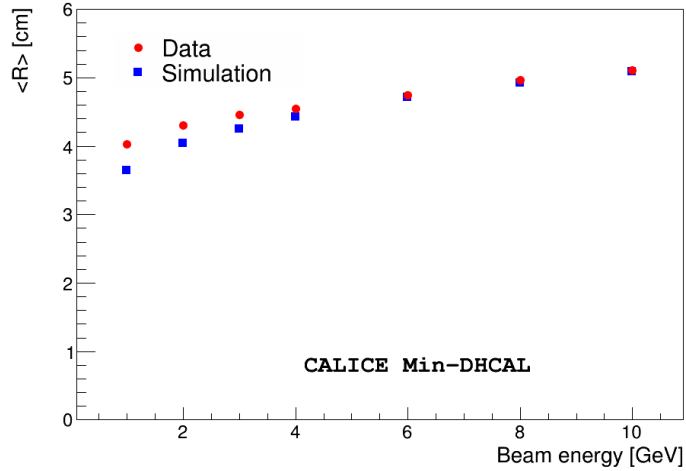

**Figure 13.** Distance of hits in an event to the fitted shower axis, averaged over all hits and events at a given beam energy: data (red) and simulation (blue). The error bars including statistical and systematic uncertainties for the data and statistical uncertainties only for the simulation are smaller than the marker size.

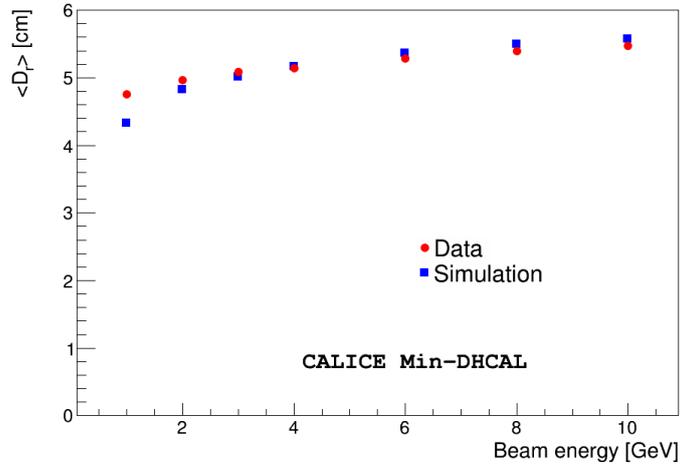

**Figure 14.** Radial dispersion of hits in an event versus beam energy: data (red) and simulation (blue). The error bars including statistical and systematic uncertainties for the data and statistical uncertainties only for the simulation are smaller than the marker size.

## 10. Linearization of the Min-DHCAL Response

The detailed spatial information available from imaging calorimeters can be used to apply corrections to the measured number of hits which might result in an improved linearity of the response and energy resolution. In order to estimate the contribution of leakage out of the back of the calorimeter to the response to positrons, the fit functions to the measured longitudinal shapes were extended beyond the actual depth of the DHCAL, see Fig. 9. Integration of the curves provides an estimate of the effect of leakage. Figure 16 shows the response (peak position of the number of hits) as a function of beam energy both before and after this average leakage correction. The effect is seen to be small and to increase up to 3% at 10 GeV.

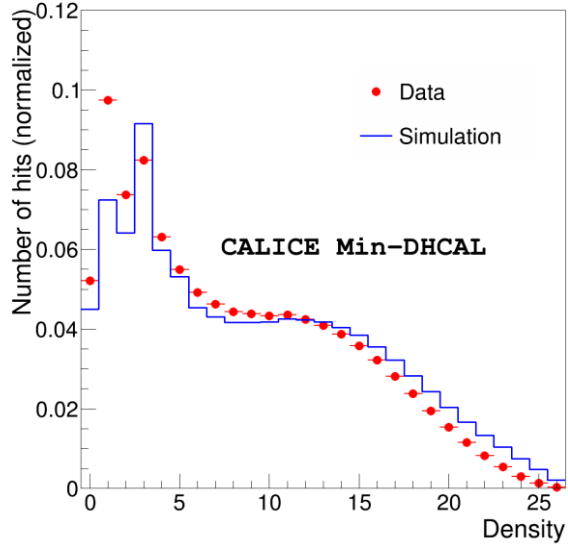

**Figure 15.** Distribution of the density of hits in events induced by 6 GeV positrons: data (red) and simulation (blue). The errors bars are significantly smaller than the plotted data points. The entries in the plot have been normalized to unity.

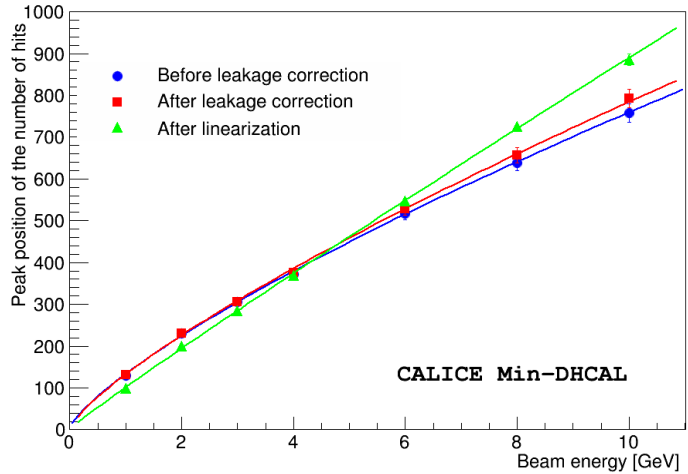

**Figure 16.** Measured peak position of the number of hits for positrons as a function of beam energy: before any correction (blue), after leakage corrections (red) and after linearization of the response based on the hit densities (green).

Next, in an attempt to improve the energy resolution for positrons, the hits from a subsample of events recorded at each of the seven energy points were weighted depending on their hit density. The weights were determined by minimizing the following quantity

$$\chi^2 = \sum_{i=1}^{7} \frac{\left(\sum_{j=0}^{26} w_j D_{ij} - \alpha E_i^{beam}\right)^2}{E_i^{beam}} \quad (9)$$

where the first sum is over the different beam energies, the $D_{ij}$'s indicate the number of hits collected at $E_i^{beam}$ with a hit density $j$, $w_j$ are the weights for hit density bin $j$, and $\alpha$ is an arbitrary scaling factor, here taken to be 90/GeV. The weights obtained by the linearization procedure are shown in Fig. 17 and are seen to be large at both low and high hit densities. The large values of the weights related to higher hit densities compensate for the saturation effects introduced by the high density of electromagnetic showers and the finite pad size of the readout boards. The large weight for the '0' density bin compensates for the unit pad multiplicity compared to the average of ~1.6. The result of this linearization procedure is also shown in Fig. 16. The parameters of the fits to power laws of the uncorrected and corrected responses are summarized in Table IV. As can be seen from the table, the linearization procedure significantly improves the linearity, but fails to achieve a perfectly linear response. Small differences between the unweighted results of Table IV and the data of Fig. 5 are due to the fact that, for technical reasons, the event samples in the linearization studies were limited to 10,000 for each energy bin.

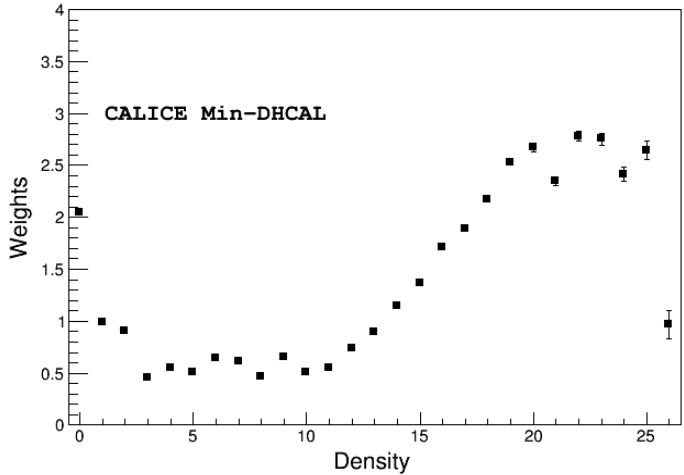

**Figure 17.** Weights as a function of hit density as obtained by the linearization procedure.

Figure 18 shows the energy resolution as a function of beam energy obtained both before and after the linearization procedure. The points have been corrected for the contribution from the known momentum spread of the test beam [15]. The density-weighted linearization procedure results in a modest improvement of about 10%, as can be seen as well from the results of the fits summarized in Table V.

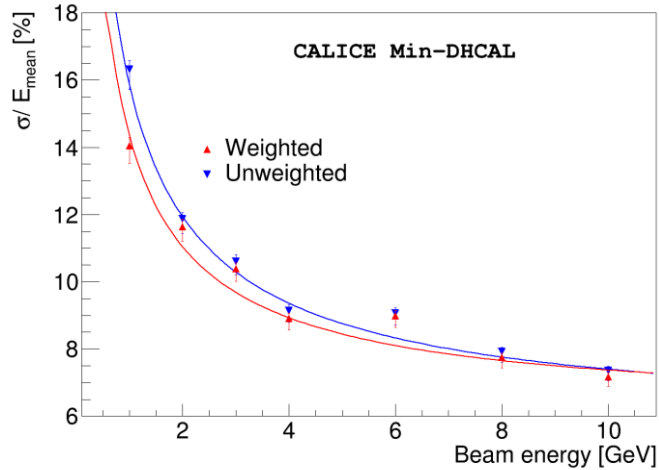

**Figure 18.** Energy resolution versus beam energy for positrons: before (blue) and after (red) the linearization procedure.

**Table IV.** Parameters obtained from the fits of the peak position of the number of hits versus beam energy to power laws $\alpha E^m$.

| Data | $a$ [GeV$^{-1}$] | $m$ |
|---|---|---|
| Before corrections | 132±3 | 0.76±0.02 |
| After leakage corrections | 133±3 | 0.78±0.02 |
| After linearization | 99±2 | 0.94±0.01 |

**Table V.** Summary of results of the fits to the energy resolution as function of energy, see Eq.4.

| Fit | $c$ [%] | $\alpha$ [%] |
|---|---|---|
| Unweighted | 5.7 ± 0.2 | 14.8 ± 0.4 |
| Weighted (linearized) | 6.2 ± 0.2 | 13.0 ± 0.4 |

## 11. Conclusions

The Digital Hadron Calorimeter (DHCAL) detector planes without absorber plates, the Min-DHCAL, was exposed to particles in the Fermilab test beam. The response of the individual Resistive Plate Chambers in the calorimeter stack was equalized using through-going muon tracks. The response of the Min-DHCAL to positrons, its energy resolution and various electromagnetic shower shapes were measured in the energy range of 1 to 10 GeV. The results of a Monte Carlo simulation based on GEANT4 and a standalone program, RPC_sim, to emulate the response of the RPCs, were compared to the data. The RPC_sim program was tuned to reproduce the measured response to muons and to reproduce the measurements obtained with 3 and 10 GeV positrons. Due to the tuning process the simulation lost its predictive power for both muons and positrons.

The GEANT4 simulation utilized both the FTFP_BERT and the FTFP_BERT_EMY physics lists. The latter provides higher accuracy, in particular for the simulation of electromagnetic processes in thin layers. Despite tedious efforts of tuning of the RPC_sim parameters to reproduce the measurements, only a poor description of the data was obtained with FTFP_BERT, suggesting a deficit of ionizations in the gas gap of the RPCs. A significant improvement is seen with the use of FTFP_BERT_EMY, leading to a good to excellent agreement with the data.

In a follow-up paper, the simulation based on the parameters obtained here with muons and positrons will be confronted with the measurements performed with pions. Since there will be no further tuning of the simulation, the results of the simulation of pion showers will retain their predictive power, albeit within systematic uncertainties related to the discrepancies observed in the analysis of the positron data.

**Acknowledgments**

The authors would like to thank the FTBF team of Fermilab, in particular Aria Soha, Charles Brown, Richard Coleman, Todd Nebel, and Erik Ramberg, for their invaluable help and great assistance. It was a major achievement to provide stable beams at much reduced rates, which were then acceptable for the rate-limited RPCs in the Min-DHCAL.

We would like to thank the technicians and the engineers who contributed to the design and construction of the prototype. This work was supported by the Bundesministerium für Bildung und Forschung (BMBF), Germany; by the Deutsche Forschungsgemeinschaft (DFG), Germany; by the Helmholtz-Gemeinschaft (HGF), Germany; by the Alexander von Humboldt Stiftung (AvH), Germany; by the Russian Ministry of Education and Science contracts 4465.2014.2 and 14.A12.31.0006 and the Russian Foundation for Basic Research grant 14-02-00873A; by MICINN and CPAN, Spain; by the US Department of Energy under contract DE-AC02-06CH11357 and the US National Science Foundation; by the Ministry of Education, Youth and Sports of the Czech Republic under the projects AV0 Z3407391, AV0 Z10100502, LG14033 and 7E12050; and by the National Sciences and Engineering Research Council of Canada.